\documentclass[11pt, twocolumn]{article} 
\usepackage[font=footnotesize]{caption}
\usepackage{subcaption}
\usepackage{graphicx} 
\usepackage{float}
\usepackage{authblk} 
\usepackage[T1]{fontenc} 

\usepackage{graphicx} 
\usepackage[textwidth=16cm,textheight=21.7cm]{geometry}

\usepackage[caption=false,font=normalsize,labelfont=sf,textfont=sf]{subfig} 

\usepackage{physics}

\usepackage{amsmath} 

\title{\textbf{Loss Tomography for Quantum Networks}}
\author[1]{Jake Navas}
\author[1]{Jaden Brewer}
\author[1]{Jaime Diaz}
\author[2]{Matheus Guedes de Andrade}
\author[2]{Don Towsley}
\author[1]{Inès Montaño}
\affil[1]{Department of Applied Physics and Materials Science, Northern Arizona University}
\affil[2]{Manning College of Information and Computer Science, University of Massachusetts Amherst}
\date{}

\begin{document}

\maketitle

\textbf{\textit{Abstract} - With steady progress in the development of quantum networks, the question on how to best provide end-to-end characterization of such networks (Quantum Network Tomography) is quickly becoming more pressing. Initial results demonstrated how we can utilize multipartite entanglement distribution to determine error probabilities of single-Pauli channels and depolarizing channels. In this work, we show how the analysis of quantum capacity regions can be used as a powerful new tool in quantum network tomography. As a first application of the proposed method, we demonstrate how we can characterize the loss on quantum channels in the network directly from quantum capacity region diagrams, even in the presence of bit-flip errors. Our results indicate that quantum capacity regions are not only valuable for network design, resource allocation, and protocol benchmarking, but also show promise for applications in quantum network tomography, particularly in loss tomography.}
\section{Introduction}
The realization of large-scale quantum networks is a crucial requirement for multiple quantum 2.0 applications such as distributed quantum computation\textsuperscript{\cite{nielsen2010quantum}}, secure communication\textsuperscript{\cite{BB84, ShorBB84}}, and highly precise quantum sensing\textsuperscript{\cite{Ent-ass-sens, sensing}}. While progress in the development of quantum networks has been steady over the last years, many obstacles still need to be overcome before large-scale quantum networks can be realized. Pressing challenges are related to present hardware limitations, such as the development of quantum repeaters. However, as hardware continues to improve and networks start to scale, the question of how to efficiently characterize and monitor the quality of quantum networks becomes increasingly pressing. Inspired by classical network tomography\textsuperscript{\cite{vardi1996network, towsley2002network}}, Quantum Network Tomography (QNT) aims to characterize errors of internal network components through end-to-end measurements\textsuperscript{\cite{QNT2022,QNT2023,QNT2024}}.  

In previous work, we demonstrated the ability of initial QNT protocols to infer error probabilities of single-Pauli channels and depolarizing channels in quantum star networks\textsuperscript{\cite{QNT2022,QNT2023,QNT2024}}, assuming error-free quantum gates and memories. The proposed protocols utilizes multipartite entanglement distribution to characterize the quality of states transmitted through the network. While Pauli error rates and depolarization probabilities can successfully be inferred for all channels of the network, these initial protocols did not address loss estimation. 

In this work, we present a novel approach to QNT built around the analysis of quantum capacity regions (QCRs), which, as we will show below, is ideally suited for loss tomography. QCRs of quantum switches have recently been studied to provide guidance on how to achieve maximum end-to-end user entanglement generation rates under given physical constraints such as noisy channel transmissions and imperfect quantum operations \textsuperscript{\cite{vasantam2022throughput,vardoyan2023capacity,panigrahy2023capacity}}. We characterize here how noise impacts the capacity region and, most importantly, demonstrate how we can utilize 2-D capacity regions to fully characterize loss on all quantum channels in a three-node star quantum network, even in the presence of bit-flip errors. 

\section{Methods}
\subsection{Model System and Network Simulation}

\begin{figure*}
    \centering
    \begin{subcaptionbox}{Topology and Routing\label{fig:subfigA}}[0.45\textwidth]
    {\raisebox{0.75cm}{\includegraphics[width=\linewidth]{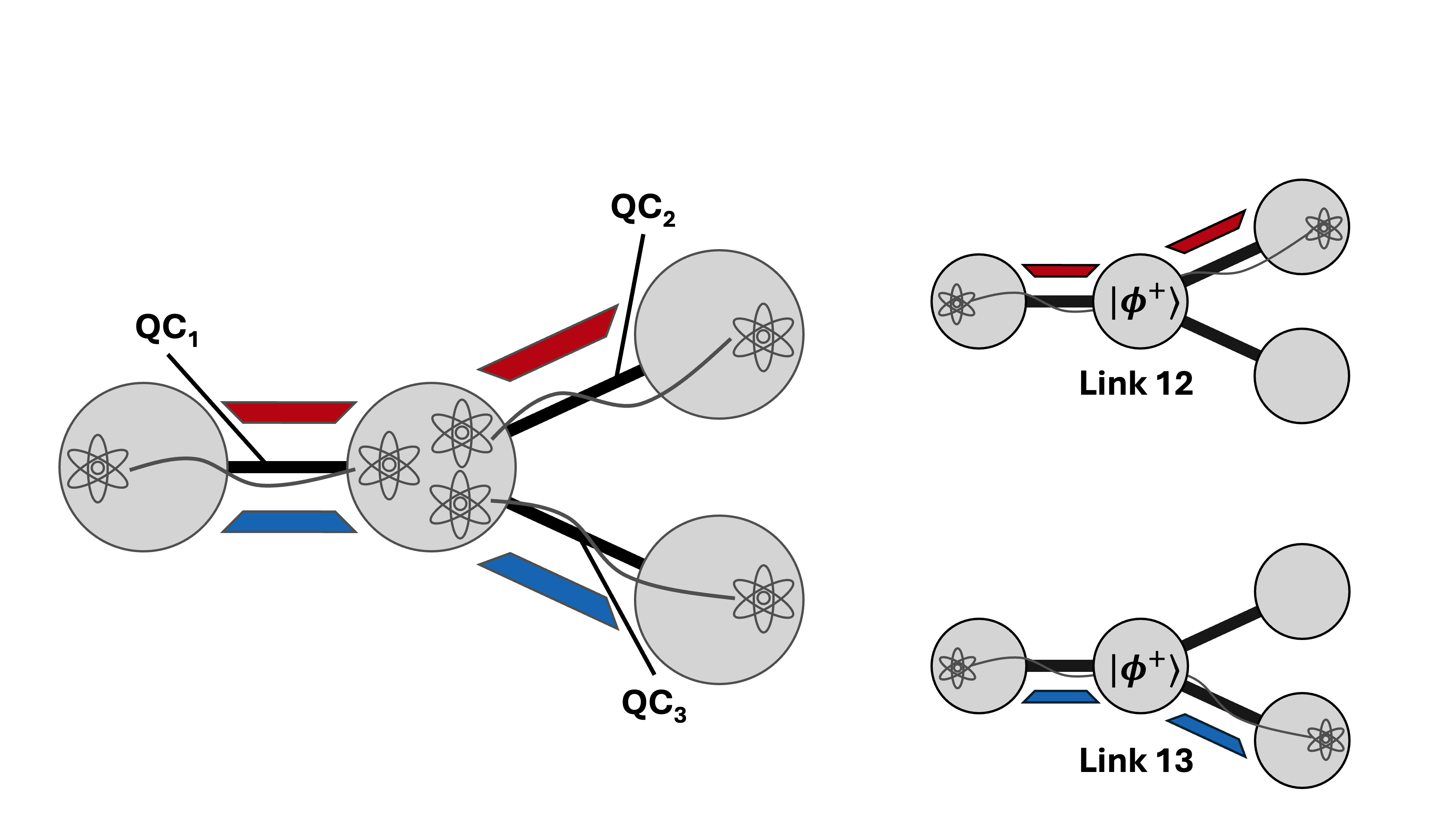}}}
  \end{subcaptionbox}
  \hspace{0.05\textwidth}
  \begin{subcaptionbox}{Generated Capacity Region\label{fig:subfigB}}[0.45\textwidth]
    {\includegraphics[width=\linewidth]{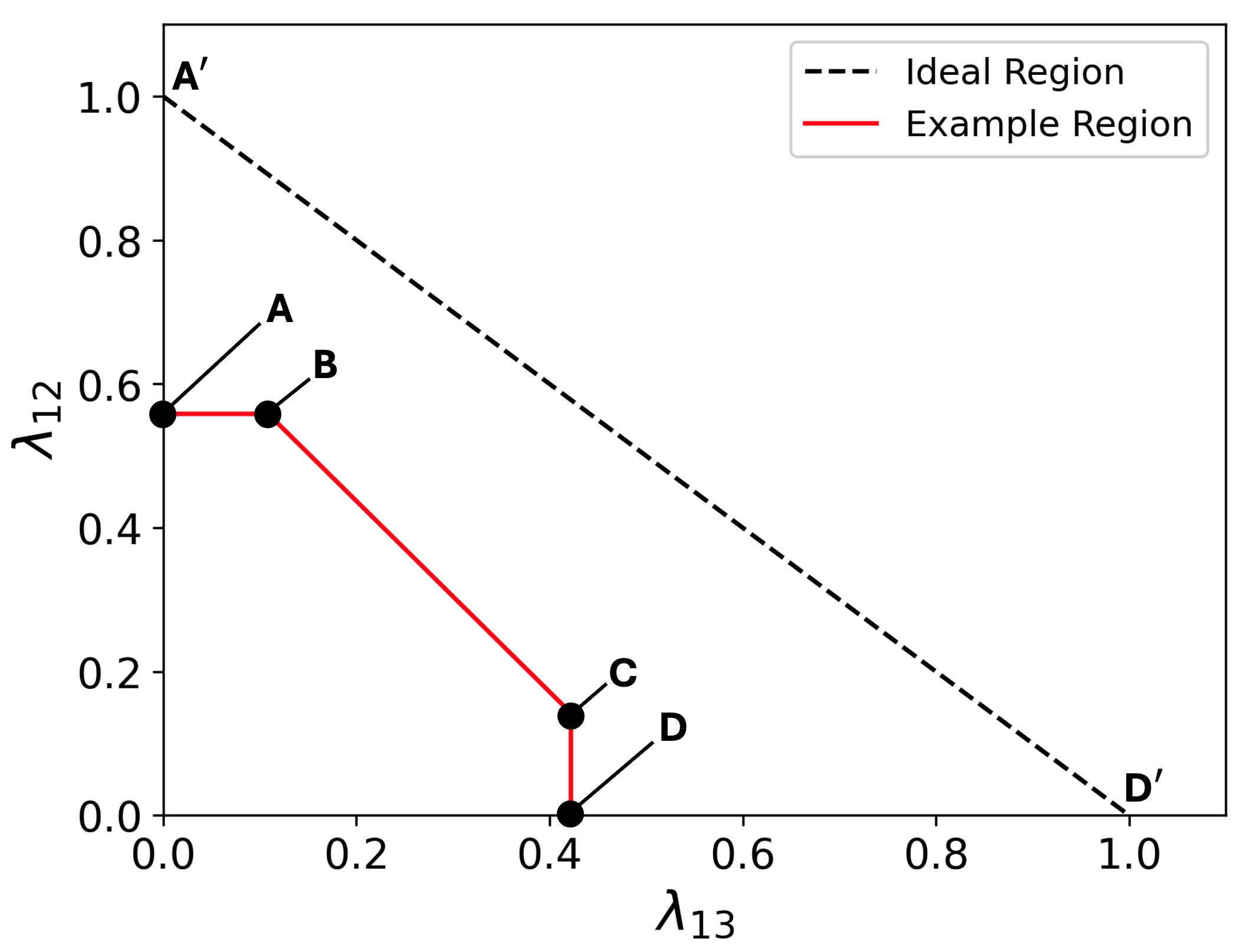}}
  \end{subcaptionbox}
  \caption{\small (a) End-to-end entanglement path generated between $N_1$ and $N_2$ in red. Repeated for $N_1$ and $N_3$ in blue (b) Example capacity regions for an ideal, noiseless network (dashed) and a network with heterogeneous loss and bit-flip noise  model (red).}
  \label{fig:mainfig}
\end{figure*}
In the following, we consider the rooted tree network shown in Figure~\ref{fig:mainfig}(a). Each of the three user nodes $(N_1$, $N_2$, $N_3)$ are connected to a central switch node (SN) via a quantum channel ($QC_1$, $QC_2$, $QC_3$) and a classical channel ($CC_1$, $CC_2$, $CC_3$). 

To provide users with end-to-end entanglement between a root node and one of two leaf nodes, each user node locally prepares a Bell pair in the state $\ket{\phi^+}=\frac{1}{\sqrt{2}}\Big(\ket{00} + \ket{11}\Big )$ and distributes one qubit from its Bell pair to the SN via the corresponding quantum channel. When the SN receives a request for end-to-end entanglement between the root node, $N_1$, and a leaf node, $N_j$, where $j=2$ or $j=3$, it performs entanglement swapping on the qubits belonging to the corresponding channels, i.e., $QC_1$ and $QC_j$.

To systematically study the impact of noisy channels on the quantum capacity region of the network and to benchmark the feasibility of the proposed loss tomography protocol, we simulate the rooted tree network described above using the discrete-event quantum network simulator NetSquid\textsuperscript{\cite{NetSquid}}. In particular, we utilize the simulated network as a virtual testbed, which allows us to explore the impact of noise on a network's QCR and to verify if our proposed loss tomography protocol correctly recovers the underlying loss model.

\subsection{Extracting Quantum Capacity Regions}
Generally, capacity regions are defined as the arrival processes that can be stably supported by a network, given a scheduling policy\textsuperscript{\cite{gupta2000capacity}}. In this work we are trying to determine if and how noise in the network impacts the capacity region. We extract the capacity region of a given network by adapting the general scheme laid out by N. K Panigrahy, et. al\textsuperscript{\cite{panigrahy2023capacity}}. They define the capacity region as the set of request rates $\{\lambda_{12},\lambda_{13}\}$ for end-to-end entanglement that can be reliably supported within given physical constraints. In our case, the given physical constraints include channel noise the system experiences as well as constraints from the scheduling policies we impose. 

We assume that during each time slot, each of the user nodes creates link-level entanglement with the SN as described above. We assume that with rate $\lambda_{1j}$, the switch receives end-to-end entanglement requests between the root node, $N_1$, and leaf node, $N_j$, respectively. To fulfill a request, the switch needs to successfully perform entanglement swapping on the qubits belonging to the corresponding links $QC_1$ and $QC_j$. We assume that the network can only fulfill one request for end-to-end entanglement per time slot, resulting in a maximum of one end-to-end entanglement per time slot. Unused link-level entanglement is discarded at the end of each time step. This entanglement distribution setup for a rooted star network yields a two-dimensional QCR.

To extract the QCR of a given network, we track the network's ability to fulfill received end-to-end entanglement requests with the following routing conditions. We consider the request arrival vector $\lambda = [\lambda_{12},\lambda_{13}]$. To determine the maximum supportable request rate, $\lambda_{1j,max}$ (where $j=2,3$), for end-to-end entanglement between the root node $N_1$ and leaf node $N_j$, we set $\lambda_{1k}=0$ (where $j\neq k$ and $k=2,3$) and record how many times the request for $N_1$-$N_2$ entanglement can be fulfilled in 10,000 independent trial runs. $\lambda_{12,max}$ is recorded as QCR reference point $A$ and $\lambda_{13,max}$ is recorded as QCR reference point $D$ in the corresponding capacity diagram (Figure~\ref{fig:mainfig}(b)). In the noiseless case (no errors, no loss) nothing prevents the switch from successfully fulfilling each submitted request and we find $\lambda_{12,max}$=$\lambda_{13,max}$=1. Connecting the two points defines the boundary of the capacity region in this idealized, noiseless case, depicted by the dashed line between points $A'$ and $D'$ in Figure~\ref{fig:mainfig} (b). Any combination of request rates lying between this boundary and the coordinate axes can be supported reliably by the corresponding network.

If the network experiences noise, some requests will not be fulfilled, which reduces $\lambda_{12,max}$ and $\lambda_{13,max}$ accordingly. In the case of single-Pauli errors, Z-basis measurements are performed on the heralded user-to-user entangled state in order to perform a parity check. This parity check allows the entanglement swap to be considered a failure of the requested end-to-end entanglement if the requested entangled state could not be created. If a qubit does not reach the SN due to loss, link-level entanglement cannot be established which also results in a failure for that time-step. An example of reduced capacity due to noise is shown by the solid line in Figure~\ref{fig:mainfig}(b). Our scheduling policy introduces the concept of a prioritized path and a secondary (backup) path. Allowing backup scheduling reveals additional achievable request rate pairings in the capacity region. QCR reference point B corresponds to small, nonzero $\lambda_{13}$ obtained solely through backup scheduling. The switch is allowed to serve a request for $N_1$-$N_3$ entanglement whenever a loss event occurs on QC2, which prevents fulfillment of the initially requested $N_1$-$N_2$ entanglement. Similarly, QCR reference point C corresponds to nonzero $\lambda_{12}$, also obtained solely from backup scheduling data. Please note that when we study request arrival vector $\lambda = [\lambda_{12},\lambda_{13}]$ at no time do we submit a request for $N_2$-$N_3$ entanglement due to the root/leaf dynamic of the tree network. Neither this restriction nor the scheduling policy used in the generation of these QCRs is required for the validity of our method in the characterization of loss. They are simplified methods of producing valid capacity regions. 
The purpose of the work presented here is to investigate the effects of particular errors on capacity regions and how we can use QCRs to infer error parameters for network links. As we will show below, QCRs allows us to infer the loss probabilities of all channels directly, even in the presence of Pauli errors.

\subsection{Channel Noise}

We assume perfect quantum gates and memories in the nodes and focus our study on the impact of a lossy, bit-flip channel, $\mathcal{E}_i$ (for $i=1,2,3) $, of the form
\begin{align}\label{Eq:1}
    \begin{split}
    \mathcal{E}_i(\rho)&=\mathcal{L}_i\Big[\theta_i\rho+(1-\theta_i)X\rho X\Big]\\
    &\hspace{1.9em}+(1-\mathcal{L}_i)\ket{l}\bra{l}\\
    \end{split}
    \end{align}

Here, $\rho$ is the input state before interacting with the quantum channel $\mathcal{E}_i$, $(1-\theta_i)$ is the probability of a bit-flip error with $0 \leq \theta_i \leq 1$, ($1-\mathcal{L}_i)$ is the probability of loss with $0 \leq \mathcal{L}_i \leq 1$, $X$ is the Pauli X operator, and $\ket{l}\bra{l}$ corresponds to an erasure state that measures orthogonally to the system\textsuperscript{\cite{Bennet_erasure}}. From (\ref{Eq:1}) we can derive analytical expressions that give the (x,y)-coordinates of the four QCR reference points (A, B, C, D) as
\begin{subequations}\label{Eq:2}
\begin{align}
  A_x &= 0 \label{Eq:2a}\\
   A_y &= \mathcal{L}_1 \mathcal{L}_2 \left[\theta_1 \theta_2 + (1-\theta_1)(1-\theta_2)\right]\label{Eq:2b}\\
    B_x &= \mathcal{L}_1 (1-\mathcal{L}_2) \mathcal{L}_3 \left[\theta_1 \theta_3 + (1-\theta_1)(1-\theta_3)\right]\label{Eq:2c}\\
     B_y &= \mathcal{L}_1 \mathcal{L}_2 \left[\theta_1 \theta_2 + (1-\theta_1)(1-\theta_2)\right]\label{Eq:2d}\\
      C_x &= \mathcal{L}_1 \mathcal{L}_3 \left[\theta_1 \theta_3 + (1-\theta_1)(1-\theta_3)\right]\label{Eq:2e}\\
       C_y &= \mathcal{L}_1 \mathcal{L}_2 (1-\mathcal{L}_3) \left[\theta_1 \theta_2 + (1-\theta_1)(1-\theta_2)\right]\label{Eq:2f}\\
        D_x &= \mathcal{L}_1 \mathcal{L}_3 \left[\theta_1 \theta_3 + (1-\theta_1)(1-\theta_3)\right]\label{Eq:2g}\\
          D_y &= 0.\label{Eq:2h}
\end{align}
\end{subequations}

 As described above, this network operates as a rooted star network where channel $QC_1$ acts as the root channel, while $QC_2$ and $QC_3$ act as leaf channels, respectively. For this network configuration, we can use the set of equations given in (\ref{Eq:2}) to determine the loss rate on both of the leaf channels in networks with homogeneous and heterogeneous loss models, even in the presence of bit-flip errors. Loss on root channel $QC_1$ can be determined if, instead of request arrival vector $\lambda = [\lambda_{12},\lambda_{13}]$, we study one of the other alternate request arrival vectors such as $\lambda_A = [\lambda_{21},\lambda_{23}]$ or $\lambda_B = [\lambda_{31},\lambda_{32}]$. In particular, we redefine $N_1$ as a leaf node, rather than the root node, which allows us to determine the loss on the associated channel utilizing $N_2$ or $N_3$ as the root node by the process described above.
\section{Results}
In the following, we demonstrate how we can infer information on loss and bit-flip errors directly from QCR diagrams. 
To systematically characterize the impact of noise on the capacity region, we first consider only a single noisy channel per network. This allows us to directly associate prominent features in the QCR diagram with specifics of the underlying noise model.  

\subsection{Single Noisy Channel}


\begin{figure*}
    \includegraphics[width=\linewidth]{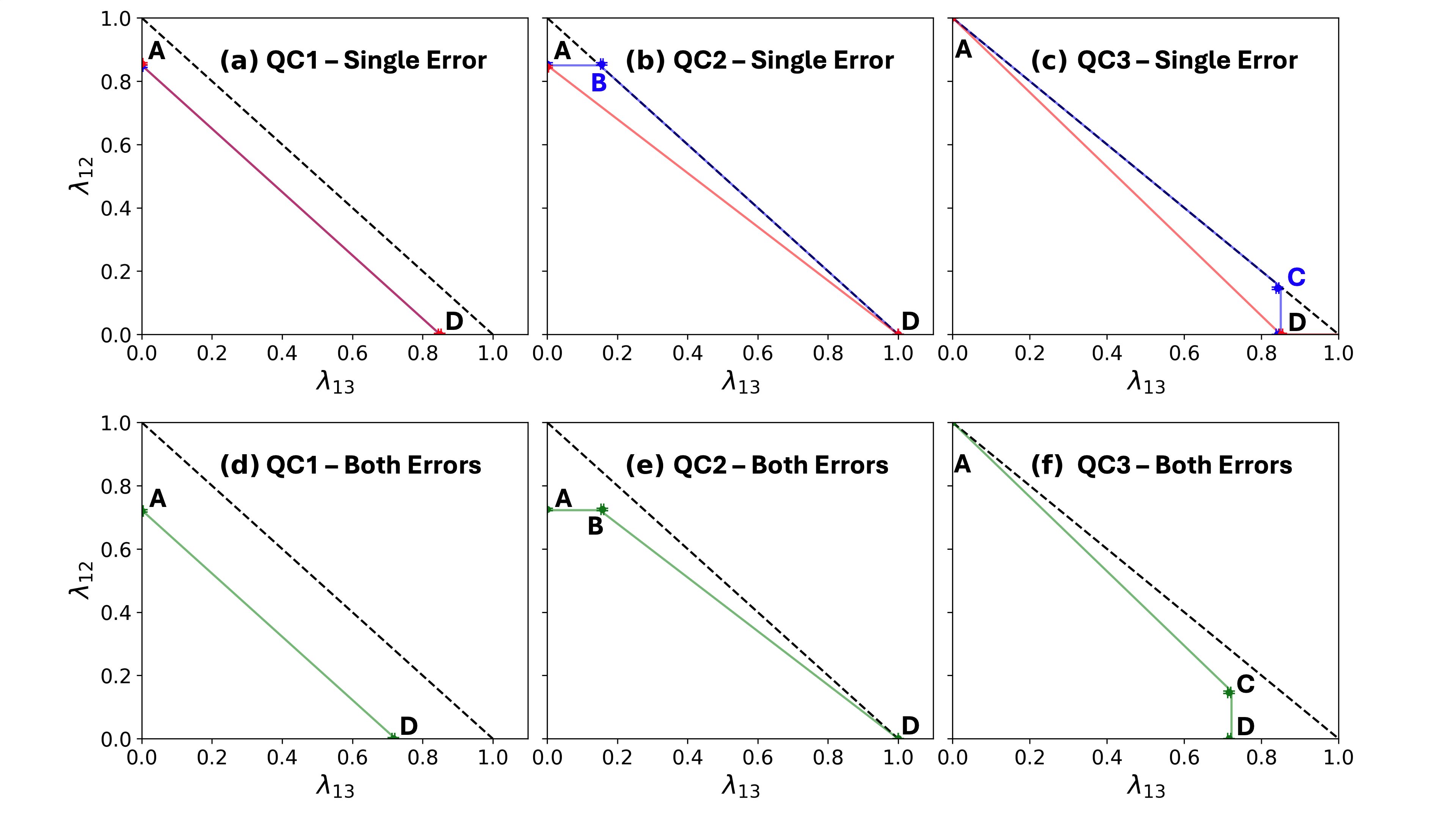}
    \caption{Dashed black indicates the ideal, noiseless network. Blue: 15\% probability of loss, red: 15\% probability of bit-flip error, green: both loss and bit-flip errors at 15\% each.
    }
    \label{fig:single channel}
\end{figure*}


Figures \ref{fig:single channel}(a), \ref{fig:single channel}(b), and \ref{fig:single channel}(c) display capacity regions for the described network when only a single quantum channel experiences noise. The dashed lines show the idealized noiseless case to facilitate comparison. Blue lines  correspond to a network experiencing a loss rate of 15\% on either channel $QC_1$ (Figure~\ref{fig:single channel}(a)), $QC_2$ (Figure~\ref{fig:single channel}(b)), or $QC_3$ (Figure~\ref{fig:single channel}(c)). Similarly, red lines correspond to a network experiencing a bit-flip Pauli rate of 15\% on the noisy channel. 

As expected, a noisy channel reduces the corresponding maximum supportable request rate(s) compared to a noiseless network. However, even without further analysis, we immediately see how the specifics of the networks are reflected in the corresponding features of the QCR. 

\subsubsection{Noisy Root Channel}
First, we discuss the behavior of a network experiencing noise on root link, $QC_1$, as shown in Figure~\ref{fig:single channel}(a). Since both possible end-to-end entanglements, $N_1$-$N_2$ and $N_1$-$N_3$, depend on the successful delivery of the qubit traveling $QC_1$, the maximum rates $\lambda_{12,max}$ and $\lambda_{13,max}$ are equally reduced (QCR points A, D). Backup scheduling cannot generate a nonzero $B_x$ or $C_y$ value, as described by (\ref{Eq:2b}) and (\ref{Eq:2f}), since backup scheduling is not possible when noise occurs only on root link $QC_1$. Consequently, QCR points B and C are not present. In this scenario, loss (blue) and bit-flips (red) affect the capacity region in the same way. In Figure~\ref{fig:single channel}(d), both error types are present with equal rates. Including both error types simultaneously changes the overall impact of noise but does not introduce new features in the capacity region. Points A and D are decreased due to the combination of error rates, while points B and C are still absent, as errors to the root still propagate evenly through both network links.

\subsubsection{Noisy Leaf Channels} A network experiencing noise only on leaf channel $QC_2$, see Figure~\ref{fig:single channel}(b), does not show the symmetric noise response observed in Figure~\ref{fig:single channel}(a) since the two types of end-to-end entanglement requests, $N_1$-$N_2$ and $N_1$-$N_3$, are not affected in the same way. While $\lambda_{12,max}$ is reduced in comparison to the ideal, noiseless case, $\lambda_{13,max}$ is unchanged since the noise present in the network does not affect this particular end-to-end entanglement. Furthermore, in contrast to Figure~\ref{fig:single channel}(a), Figure~\ref{fig:single channel}(b) shows that the impact on the capacity region depends on the noise-type. Loss and bit-flip errors manifest differently in the corresponding capacity diagram. It is only when channel $QC_2$ experiences loss that backup scheduling produces a nonzero $B_x$ value, 
\begin{align}
    B_{x_{blue}} = .1555\pm.0036,
\end{align}
described by (\ref{Eq:2b}). Using backup scheduling, the maximum supportable request rate for the network increases as the switch is allowed to utilize the $N_1$-$N_3$ path rather than discarding all qubits when the prioritized path fails. This presents as the plateau seen when $0<\lambda_{13}\leq B_x$. When $\lambda_{13}\geq B_x$, the network can reliably support any combination of request rates as a consistent ratio, following the behavior in the noiseless case. When channel $QC_2$ experiences bit-flips, QCR point B is not present. This is due to the fact that bit-flip errors are only caught in the final fidelity check, which is performed after the entanglement swapping procedure has been performed. This means bit-flip noise reduces $\lambda_{12,max}$ in a way that cannot be mitigated by backup scheduling, as the noisy path is unknowingly selected by the SN due to the qubit being present at the time of the swap. However, when a qubit is lost on $QC_2$ before entanglement swapping is attempted, backup scheduling allows the SN to instead serve an $N_1$-$N_3$ entanglement request. 

When both error types are present, we again observe a nonzero $B_x$ value on the QCR corresponding to $QC_2$ shown in Figure \ref{fig:single channel}(f). Since bit-flip errors do not offer the potential to increase throughput with backup scheduling, the benefit of allowing backup scheduling is reduced. In this case, the network can no longer reliably support any combination of request rates  with $\lambda_{13}>B_x$ in the same way as in the ideal, noiseless network. Instead, the capacity region now converges more slowly to that of the ideal case. Due to the geometric symmetry of our network, the response to noise on link $QC_3$ is a mirrored version of the response to noise on link $QC_2$.

These results are amplified in the combined systems described in Figures~\ref{fig:single channel}(e)-\ref{fig:single channel}(f). Figure~\ref{fig:single channel}(e) illustrates the decrease across the $N_1$-$N_2$ path as point A has decreased from the previous case. Notably, the plateau described by (A-B) is found to be 
\begin{align}
    B_{x_{green}} = .1576\pm.0045,
\end{align}
which is comparable to the previous case only including loss. This is indicative of how this point correlates with loss rather than the additional Pauli noise. The $N_1$-$N_3$ path remains unaffected, as expected. Noise on $QC_3$ presents similarly due to its connection to leaf node $N_3$.

\subsection{Homogeneous/Heterogeneous Noise Models}
\begin{figure}[t]
\centerline{\includegraphics[scale=.11]{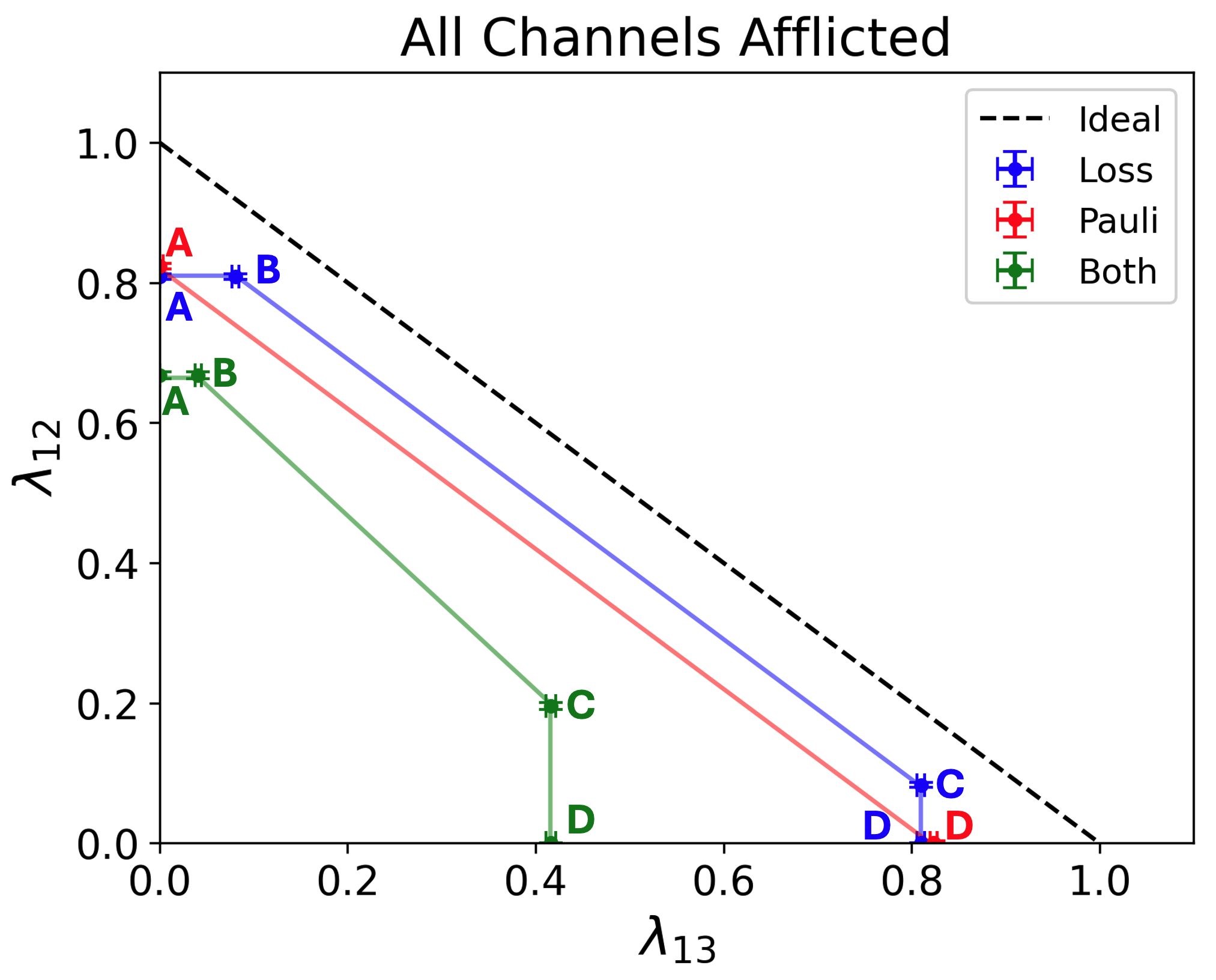}}
 \caption{Dashed black indicates the ideal, noiseless network. Blue: all channels have 15\% probability of loss, red: all channels have 10\% probability of bit-flip error, green: $QC_1$ and $QC_2$ have 10\% loss and 10\% bit-flip, whereas $QC_3$ has 30\% loss and 30\% bit-flip}
\vspace*{-5pt}
\label{fig:loss-pauli-figure}
\end{figure}
 Lastly, we investigate how the capacity region of a network changes when all quantum channels are affected by noise. As seen in Figure~\ref{fig:loss-pauli-figure}, we plot for comparison the ideal, noiseless case as well. In blue, we show the capacity region for a network with a homogeneous loss model where all three quantum channels have loss rates of 10\%, i.e. 
\begin{align}
(1-\mathcal{L}_1)=(1-\mathcal{L}_2)=(1-\mathcal{L}_3)=0.1.
\end{align} 
 In red, we show the capacity region for a network with a homogeneous bit-flip model where all three quantum channels have bit-flip rates of 10\%, i.e. 
\begin{align}
(1-\theta_1)&=(1-\theta_2)=(1-\theta_3)=0.1.
\end{align}
 Finally, in green, we show the capacity region for a network with a heterogeneous loss and bit-flip model, where $QC_1$ and $QC_2$ each have a loss rate of 10\% and bit-flip rate of 10\%, while $QC_3$ has a loss rate of 30\% and bit-flip rate of 30\%, i.e. 
\begin{subequations}
\begin{align}
(1-\mathcal{L}_1)&=(1-\theta_1)=0.1\\
(1-\mathcal{L}_2)&=(1-\theta_2)=0.1\\
(1-\mathcal{L}_3)&=(1-\theta_3)=0.3.
\end{align}
\end{subequations}
Comparing the capacity diagrams, the positive impact to the overall capacity due to backup scheduling is again visible whenever loss occurs on leaf links within the network. The plateau (A-B) shows the impact of allowing backup scheduling when requests for $N_1$-$N_2$ entanglement cannot be fulfilled due to loss on channel $QC_2$. The vertical rise (C-D) shows the impact of allowing backup scheduling when requests for $N_1$-$N_3$ entanglement cannot be fulfilled due to loss on channel $QC_3$. Examining our equations from (\ref{Eq:2}) we find that we can express the loss rate for  $QC_2$ and $QC_3$ in terms of the coordinates of the four QCR reference points:
\begin{align}
    (1-\mathcal{L}_2) &=  \frac{B_x}{D_x}\label{eq:loss}\\
    (1-\mathcal{L}_3) &=  \frac{C_y}{A_y}.\label{eq:loss2}
\end{align}
\noindent This relationship between the loss parameter and the reference points allows us to directly infer the loss rates on channels $QC_2$ and $QC_3$ from the extracted capacity region. Solving (\ref{eq:loss}) and (\ref{eq:loss2}), we find
\begin{align}
 (1-\mathcal{L}_2) &= 0.0985 \pm .0084 \\
 (1-\mathcal{L}_3) &= 0.2925 \pm .0076, 
\end{align}
\noindent which satisfactorily recovers the underlying loss model, even when bit-flip noise is also present in the network (Figure~\ref{fig:loss-pauli-figure}, green). 

We can also apply this approach to characterize the loss in the loss-only network (Figure~\ref{fig:loss-pauli-figure}, blue). Here, we determine 
\begin{align}
 (1-\mathcal{L}_2) &= 0.0994 \pm .0045 \\
 (1-\mathcal{L}_3) &= 0.1022 \pm .0045. 
\end{align} 

As discussed earlier, the impact of loss on channel $QC_1$ cannot be mitigated through backup scheduling. In contrast to the leaf channels, it is not possible to directly correlate the appearance of a plateau or vertical rise to loss on the root channel and infer its loss probability using (\ref{Eq:2}). 
To characterize the loss on root channel $QC_1$, we can modify the protocol and study one of the other alternate request arrival vectors, $\lambda_A = [\lambda_{21},\lambda_{23}]$ or $\lambda_B = [\lambda_{31},\lambda_{32}]$, instead of request arrival vector $\lambda = [\lambda_{12},\lambda_{13}]$. This allows $QC_1$ to act as one of the leaf links and, consequently, we can use (\ref{Eq:2}) to then infer the loss in this channel.

\section{Discussion}
We have shown that analyzing the capacity region of a network allows us to infer the loss probabilities on all quantum channels, even in the presence of bit-flip errors. Although not shown explicitly here, the ability to detect loss comes from the nature of the capacity region itself, rather than the method of generating the region. It is a property that is present in the QCR of any network that experiences loss, and thus, can be applied to any valid representation of a capacity region. Multiple 2-D capacity regions can be generated with different roots (differently rooted tree networks) in order to fully characterize loss across all three channels, thus allowing us to characterize loss rates for the entire system. 3-D capacity regions can also provide this capability, though in these simple cases, utilizing 2-D capacity regions is the more straightforward approach. This may not remain the case as the system scales, however, the described method is applicable to an $n$-node star network topology, making it possible to fully characterize loss in such systems. This suggests that applying these methods to larger network topologies should also be possible, but remains a subject for future work. Single-Pauli errors were not found to impact our ability to detect loss in this system, and we suspect to find similar results for other Pauli errors as well. For single-Pauli errors in particular, we did find a connection between distinct rates on network leafs and the change to the slope of the generated capacity region, but meaningful results require additional study into these effects.
\section{Conclusion}
We have characterized how noise impacts the capacity region of a network. Our results show that loss errors uniquely impact the capacity region, and that loss parameters of quantum channels can be inferred directly from the corresponding QCR. Most importantly, we demonstrated how the analysis of QCRs can be utilized to infer loss on every channel of the network, even if bit-flip errors are also present. These results open the door for analyzing larger topologies and more complicated network conditions, which makes the presented method very relevant for the field of quantum networks.
\section{Acknowledgments}
This research was supported in part by the NSF-ERC Center for Quantum Networks grant EEC-1941583.


\end{document}